\documentclass[12pt]{iopart}

\usepackage{graphicx}
\usepackage[breaklinks=true,colorlinks=true,linkcolor=blue,urlcolor=blue,citecolor=blue]{hyperref}
\begin{document}
\title[]{A nonlinearity in permanent-magnet systems used in watt balances}
\author{Shisong Li$^{1,2*}$, Stephan Schlamminger$^{2*}$, Jon Pratt$^2$}
\address{1. Department of Electrical Engineering, Tsinghua University, Beijing 100084, China\\
2. National Institute of Standards and Technology (NIST), 100 Bureau Dive Stop 8171, Gaithersburg, MD 20899-8171, USA}
\ead{leeshisong@sina.com; stephan.schlamminger@nist.gov}
\begin{abstract}
Watt balances are used to measure the Planck constant and will be used in the future to realize mass at the kilogram level. They increasingly rely on permanent magnet systems to generate the magnetic flux. It has been known that the weighing current might effect the magnetization state of the permanent magnetic system used in these systems causing a systematic bias that can lead to an error in the result if not accounted for. In this article a simple model explaining the effect of the weighing current on the yoke of the magnet is developed. This model leads to a nonlinear dependence of the magnetic flux density in the gap that is proportional to the squared value of the coil current. The effect arises from changing the reluctance of the yoke by the additional field produced by the coil. Our analysis shows that the effect depends on the width of the air gap, the magnetic flux density in the air gap, and the $BH$ curve of the yoke material. Suggestions to reduce the nonlinear effect are discussed.
\end{abstract}


\maketitle

\section{Introduction}

The watt balance, originally proposed by B. P. Kibble in 1975~\cite{kibble75}, is an instrument that is used to precisely measure the Planck constant $h$. In the new International System of Units (SI)~\cite{mills06} it will constitute one method to realize the unit of mass at the kilogram level. Currently, several national metrology institutes (NMIs) are in the process of building a watt balance, since it is seen as an ideal apparatus to realize and maintain the unit of mass in the new SI. A review on watt balance experiments is given in~\cite{steiner13}.

Typically, the watt balance is operated alternately in two separated modes: In weighing mode, a magnetic force is generated by passing a dc current $I$ through a coil in an area with a magnetic flux density $B$. The magnetic force is balanced by the gravitational force acting on a test mass $m$, i.e., $BLI=mg$, where $L$ is the wire length in the coil and $g$ the gravitational acceleration; in velocity mode, the geometric factor $BL$ is calibrated by moving the coil in the same magnetic field with a velocity $v$ while measuring an induced voltage $U$ across the coil, i.e., $U=BLv$. The combination of the two measurements allows a direct comparison of electrical power to mechanical power. The Planck constant can be obtained since electrical power can be measured as the product of two frequencies and $h$ by the virtue of the Josephson effect~\cite{josephson1962} and the quantum Hall effect~\cite{klitzing80}.

For the watt balance experiment to work, it is assumed that $BL$ is the same in two modes. However, in weighing mode the current in the coil produces a magnetic field that could cause a change in the magnetic flux density $B$, i.e., the magnetic flux $B$ is in general a function of the weighing current $I$ expressed in Taylor series \cite{lan07}:
\begin{equation}
(BL)_\mathrm{w}\approx(BL)_\mathrm{v}(1+\alpha I+\beta I^2),
\label{eq_nonlinear}
\end{equation}
where $(BL)_\mathrm{w}$ and $(BL)_\mathrm{v}$ are the geometric factors in weighing mode and velocity mode, respectively. $\alpha$ and $\beta$ denote the linear and quadratic coefficients. The weighing mode is typically carried out in a fashion that the linear term is eliminated: Two measurements, mass-off and mass-on, are performed during weighing mode \cite{NIST14}. The currents in mass-off and mass-on measurements are equal and opposite, canceling any effect caused by $\alpha$. The quadratic term, however, cannot be eliminated and can lead to a bias in the measurement.

The nonlinear effect caused by the parallel component of the weighing flux has been studied in~\cite{li13} and $\beta$ was determined by considering the magnetic reluctance change in upper and lower yokes. It was found that the main part of the nonlinear error from the parallel component is canceled by averaging the upper and lower yokes. In the end, the size of the bias in the measurement introduced by this component is negligible compared to the desired accuracy of the watt balance, which is typically a few parts in $10^{8}$. Recently, a different mechanism that can produce a quadratic dependence of $BL$ on the current was found while investigating the NIST-4 magnet~\cite{schlamminger13,seiffert14} at National Institute of Standards and Technology, USA. The quadratic term arises due to a change in reluctance of the yoke near the coil caused by the perpendicular component of the additional magnetic field $H$ created by the weighing current. 
In this article we demonstrate the origin of this nonlinear effect, estimate its order of magnitude, and discuss strategies to reduce or even remove this error by design improvements, active compensation, or corrections.

\begin{figure}[!tp]
\centering
\includegraphics[width=3.7in]{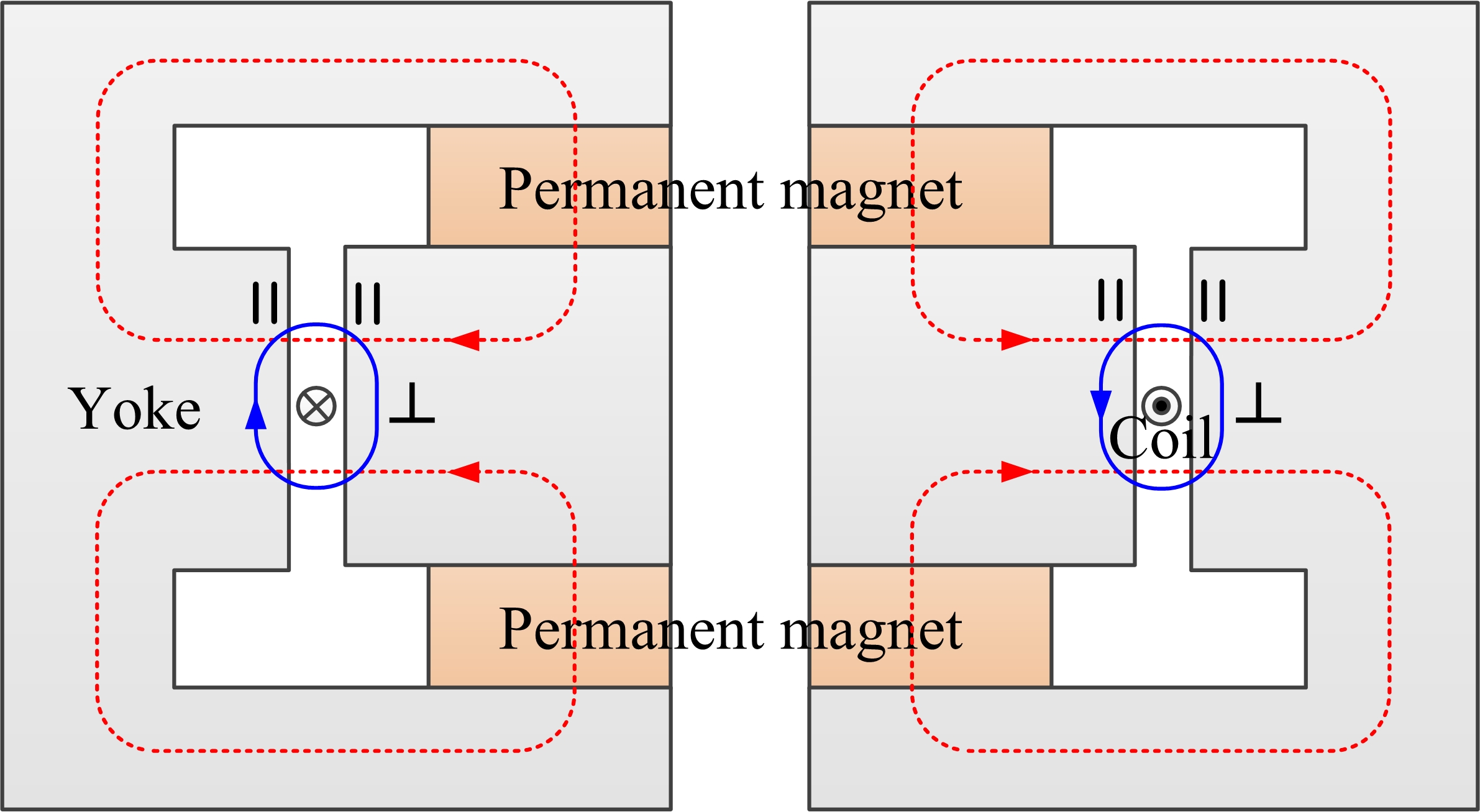}\\
(a)\\
~\\
\includegraphics[width=3.7in]{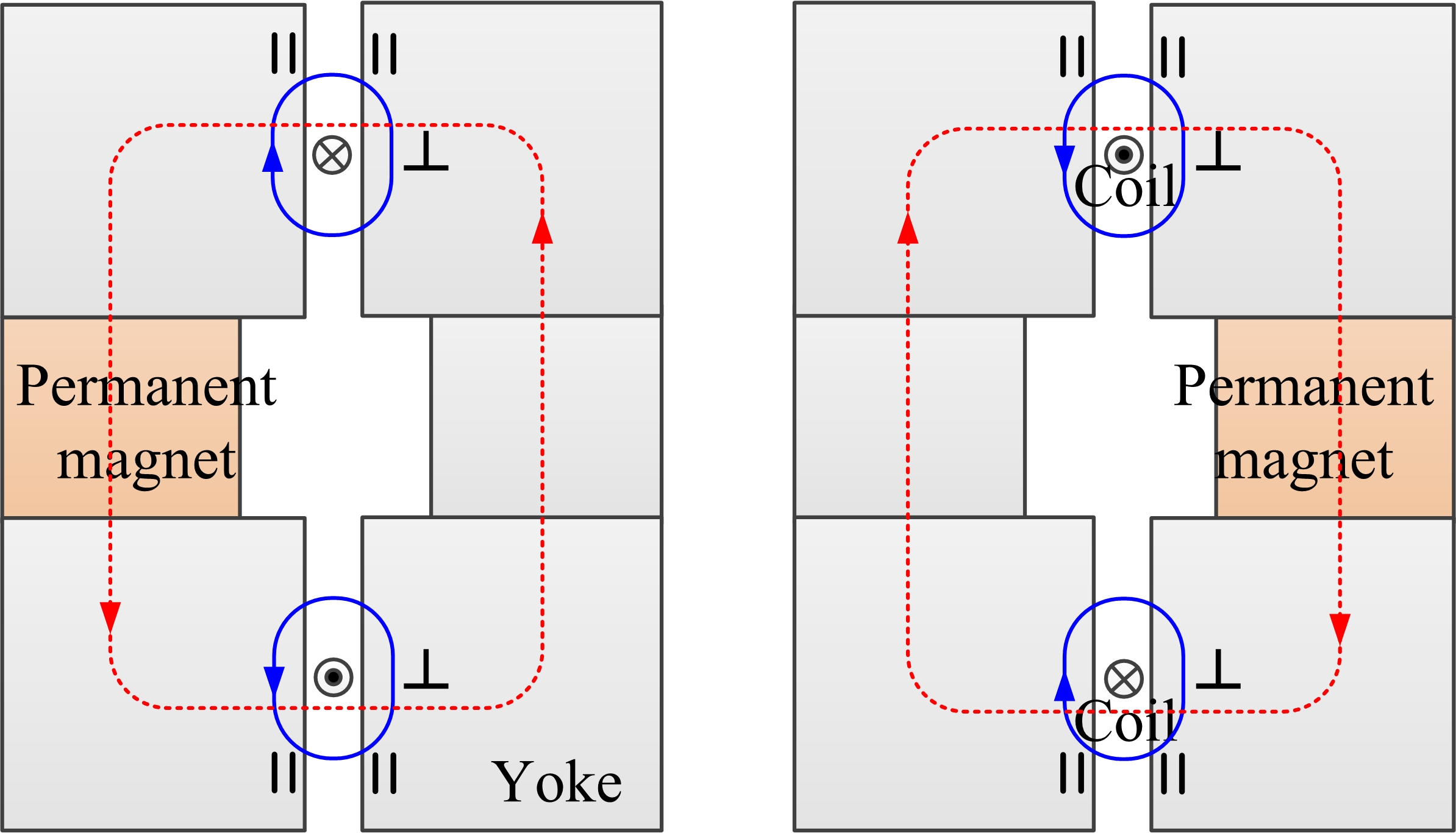}\\
(b)\\
\caption{Typical magnetic circuits employed in watt balance experiments. (a) Two-permanent-magnets, one-coil structure; (b) one-permanent-magnet, two-coils structure. The magnets exhibit cylindrical and up-down symmetry, where O denotes the geometric center. The red dashed line indicates the main magnetic flux generated by the permanent magnet(s); the blue line shows the additional magnetic field produced by the coil in weighing mode. The symbols $\parallel$ and $\perp$ denote the locations, where the field from the coil is mostly parallel and perpendicular to the flux from the permanent magnet(s).}
\label{mag_structure}
\end{figure}

\section{Magnetic error analysis}

Recently, yoke-based permanent-magnet systems seem to be the preferred choice in watt balances~\cite{schlamminger13,stock06,baumann13,robinson12,gournay05}. Compared to electro magnets, these systems benefit from a stronger magnetic field, lower operating cost, and better magnetic self-shielding. Figure \ref{mag_structure} shows two typical designs for such magnet systems. The two-permanent-magnet, one-coil structure as shown in figure \ref{mag_structure}(a) is employed by the BIPM watt balance~\cite{stock06} at the Bureau International des Poids et Mesures, METAS-2~\cite{baumann13} at the Federal Institute of Metrology, Switzerland, and NIST-4~\cite{schlamminger13}. The one-permanent-magnet, two-coil structure as shown in figure \ref{mag_structure}(b) is built into the NPL-NRC watt balance at the National Physical Laboratory, UK and the National Research Council, Canada~\cite{robinson12,steele12}. In this article, we focus our analysis on the two-permanent-magnet, one-coil structure. However, our results can also applied to the one-permanent-magnet, two-coil case.

Two different, but related, effects of the weighing current on the magnetic flux in the gap aspects are investigated: First, the total flux through the gap varies due to a change of the reluctance of the yoke. Second, the flux redistributes around the coil in the air gap since the reluctance of the iron closer to the coil changes differently than that of the iron farther away from the coil.

\subsection{Total magnetic flux change}
\label{2.1}

In the magnetic circuits shown in figure \ref{mag_structure}, the main magnetic flux runs horizontally through a small air gap in the yoke. In weighing mode, the coil generates an additional magnetic field, and part of this field will penetrate the yoke. As a result, the magnetic reluctance in some areas of the yoke will change, and hence alter the reluctance of the complete magnetic circuit. Therefore, the $BL$ in weighing mode will slightly differ from its value in velocity mode.

In velocity mode, the magnetic equation of the  magnet circuit can be written as
\begin{equation}
\mathcal{F} = R_\mathrm{v}\phi_\mathrm{v} \;\mbox{with}\;R_\mathrm{v}=(\frac{l_0}{\mu_0S_0}+\frac{l_\mathrm{m}}{\mu_\mathrm{m} S_\mathrm{m}}+\frac{l_\mathrm{y}}{\mu_\mathrm{y}S_\mathrm{y}}),
\label{eq_velocity}
\end{equation}
where $R_\mathrm{v}$  denotes the reluctance of the magnetic circuit in velocity mode, $\phi_\mathrm{v}$ the flux through the magnetic system and $\mathcal{F}$ the magnetomotive force of the permanent magnets. As shown on the right of equation (\ref{eq_velocity}), the total reluctance of the system is a sum of three parts: the reluctances of the air gap, the permanent magnet and the yoke. Here, $l_0$, $l_\mathrm{m}$, $l_\mathrm{y}$ denote the magnetic reluctance lengths, $S_0$, $S_\mathrm{m}$, $S_\mathrm{y}$ the magnetic reluctance areas, and $\mu_0$, $\mu_\mathrm{m}$, $\mu_\mathrm{y}$ the permeability of the air gap, the permanent magnets and the yoke. The reluctances of various magnetic paths depend on the exact geometries, which can be difficult to evaluate. In this article, all values for the areas and lengths of different flux paths are equivalent, i.e., average, values. In equation (\ref{eq_velocity}), $\mu_\mathrm{m}\approx\mu_0<<\mu_\mathrm{y}$, thus the total magnetic reluctance is dominated by the contributions of the permanent magnet and the air gap.

In weighing mode, the current in the coil generates additional fields in the yoke. The additional fields are separated into two components: parallel (subscript $_\parallel$) and  perpendicular (subscript $_\perp$) to the flux generated by the permanent magnet. The magnetic equation in weighing mode can be expressed as
\begin{equation}
\mathcal{F} = R_\mathrm{w}\phi_\mathrm{w} \;\mbox{with}\;R_\mathrm{w}=
(\frac{l_0}{\mu_0S_0}+\frac{l_m}{\mu_mS_m}+\frac{l_\parallel}{\mu_{\mathrm{w}\parallel} S_\parallel}+\frac{l_\perp}{\mu_{\mathrm{w}\perp} S_\perp}),
\label{eq_weighing}
\end{equation}
where $l_\parallel$ and $l_\perp$ denote the magnetic reluctance length; $S_\parallel$ and $S_\perp$ the magnetic reluctance areas; and  $\mu_{\mathrm{w}\parallel}$ and $\mu_{\mathrm{w}\perp}$ is the permeability of the regions of the yoke where the field generated by the weighing current is parallel and perpendicular to the original magnetic flux. From equations (\ref{eq_velocity}) and (\ref{eq_weighing}), the relative magnetic field change can be calculated as
\begin{equation}
\frac{\phi_\mathrm{w}}{\phi_\mathrm{v}}-1=\frac{R_\mathrm{v}}{R_\mathrm{w}}-1=\frac{R_\mathrm{v}-R_\mathrm{w}}{R_\mathrm{w}}\approx \frac{R_\mathrm{v}-R_\mathrm{w}}{R_\mathrm{v}} .
\label{eq_xi1}
\end{equation}
In the last approximation $R_\mathrm{w}$ in the denominator was replaced by $R_\mathrm{v}$, since these two terms differ very little from each other.

In the three equations above, it is tacitly assumed that the magnetomotive force is independent of the current in the coil, i.e., $\mathcal{F}=\mathcal{F}_\mathrm{v}=\mathcal{F}_\mathrm{w}$. In reality, this is not the case, since the magnetic field produced by the coil during weighing mode will change the working point of the permanent magnet along the demagnetization curve. However, this effect depends linearly on the weighing current and will cancel by current reversal (mass-on vs. mass-off).

To simplify the analysis, we split the reluctance of the yoke during velocity mode in the same two regions as in the weighing mode, yielding
\begin{equation}
\frac{l_\mathrm{y}}{\mu_\mathrm{y}S_\mathrm{y}}=\frac{l_\parallel}{\mu_{\mathrm{v}\parallel}S_\parallel}+\frac{l_\perp}{\mu_{\mathrm{v}\perp}S_\perp},
\end{equation}
$\mu_{\mathrm{v}\parallel}$ and $\mu_{\mathrm{v}\perp}$ are the permeabilities of two regions in velocity mode. Since there is no current in the coil during velocity mode, the symbols $\parallel$ and $\perp$ only denote the yoke locations. As shown in figure \ref{mag_structure}, a watt balance magnet typically exhibits up-down symmetry. Hence the parallel component of the magnetic field of the coil will increase the field in one half of the yoke and decrease the field in the other half by the same amount, $\Delta H_\parallel$. In a small range of the yoke $BH$ curve, the $\mu_\mathrm{y}(H)$ function can be considered to be linear, leading to
\begin{equation}
\frac{l_\parallel/2}{(\mu_{\mathrm{v}\parallel}+\chi\Delta H_\parallel)S_\parallel}+\frac{l_\parallel/2}{(\mu_{\mathrm{v}\parallel}-\chi\Delta H_\parallel)S_\parallel}\approx\frac{l_\parallel}{\mu_{\mathrm{v}\parallel}S_\parallel}
\label{eq_parallel}
\end{equation}
Here, $\chi$ is the derivative of $\mu(H)$ with respect to $H$ at the working point of the yoke, i.e. $\chi=\left.\partial\mu/\partial H\right|_{H=H_\mathrm{v}}$. Equation (\ref{eq_parallel}) shows that the reluctance of the yoke parts, at which the field from the weighing current is parallel to the flux from the permanent magnet does not change between weighing mode and velocity mode in a symmetric structure. This is because the two components cancel each other. The higher order terms in equation~(\ref{eq_parallel}) are negligible compared to the watt balance uncertainty goal~\cite{li13}.

The areas of the yoke, where the field from the weighing current is perpendicular to the flux from the permanent magnets are located around the coil. In these areas, the field generated by the weighing current is much larger than in the areas where the field is parallel to the flux. In addition the cross sections of the former areas are smaller than those of the latter areas.
The magnetic field strength increases from $H_\mathrm{v}$ in velocity mode to $H_\mathrm{w}$ in weighing mode by
\begin{equation}
H_\mathrm{w}^2=H_\mathrm{v}^2+\left(\Delta H_{\perp}\right)^2 \Longrightarrow H_\mathrm{w} \approx H_\mathrm{v} + \frac{\displaystyle \left(\Delta H_{\perp}\right)^2}{2 H_\mathrm{v}}
\label{eq_Hsquare}
\end{equation}
where $\Delta H_{\perp}$ is the increment of the magnetic field strength due to the perpendicular component of the field produced by the coil. The permeability in this area is given by
\begin{equation}
\mu_{\mathrm{w}\perp} =
\mu_{\mathrm{v}\perp} + \frac{\displaystyle \left(\Delta H_{\perp}\right)^2}{2 H_\mathrm{v}} \left. \frac{\partial \mu}{\partial H} \right|_{H=H_\mathrm{v}}
\label{eq:mup}
\end{equation}

It can be seen from equation (\ref{eq_Hsquare}) that the magnetic field would increase independent of the current direction. Combining (\ref{eq_weighing}), (\ref{eq_parallel}), and (\ref{eq:mup}) allows one to rewrite (\ref{eq_xi1}) as
\begin{equation}
\frac{\phi_\mathrm{w}}{\phi_\mathrm{v}}-1 \approx  \frac{\frac{\displaystyle l_\perp}{\displaystyle \mu_{\mathrm{v}\perp}S_\perp} \left(1-\frac{\mu_{\mathrm{v}\perp}}{\mu_{\mathrm{w}\perp}}\right)}{R_\mathrm{v}}\approx
\frac{\frac{\displaystyle l_\perp}{\displaystyle \mu_{\mathrm{v}\perp}^2 S_\perp}}{R_\mathrm{v}}\frac{\displaystyle \left(\Delta H_{\perp}\right)^2}{2 H_\mathrm{v}} \left. \frac{\partial \mu}{\partial H} \right|_{H=H_\mathrm{v}}.
\label{eq_xi1F}
\end{equation}

In this section, it was assumed that the relative distribution of the flux in the air gap remains the same, i.e., is independent of the weighing current. In the next section the effects of a flux redistribution in the air gap is considered.

\begin{figure}[!tp]
\centering
\includegraphics[width=3.5in]{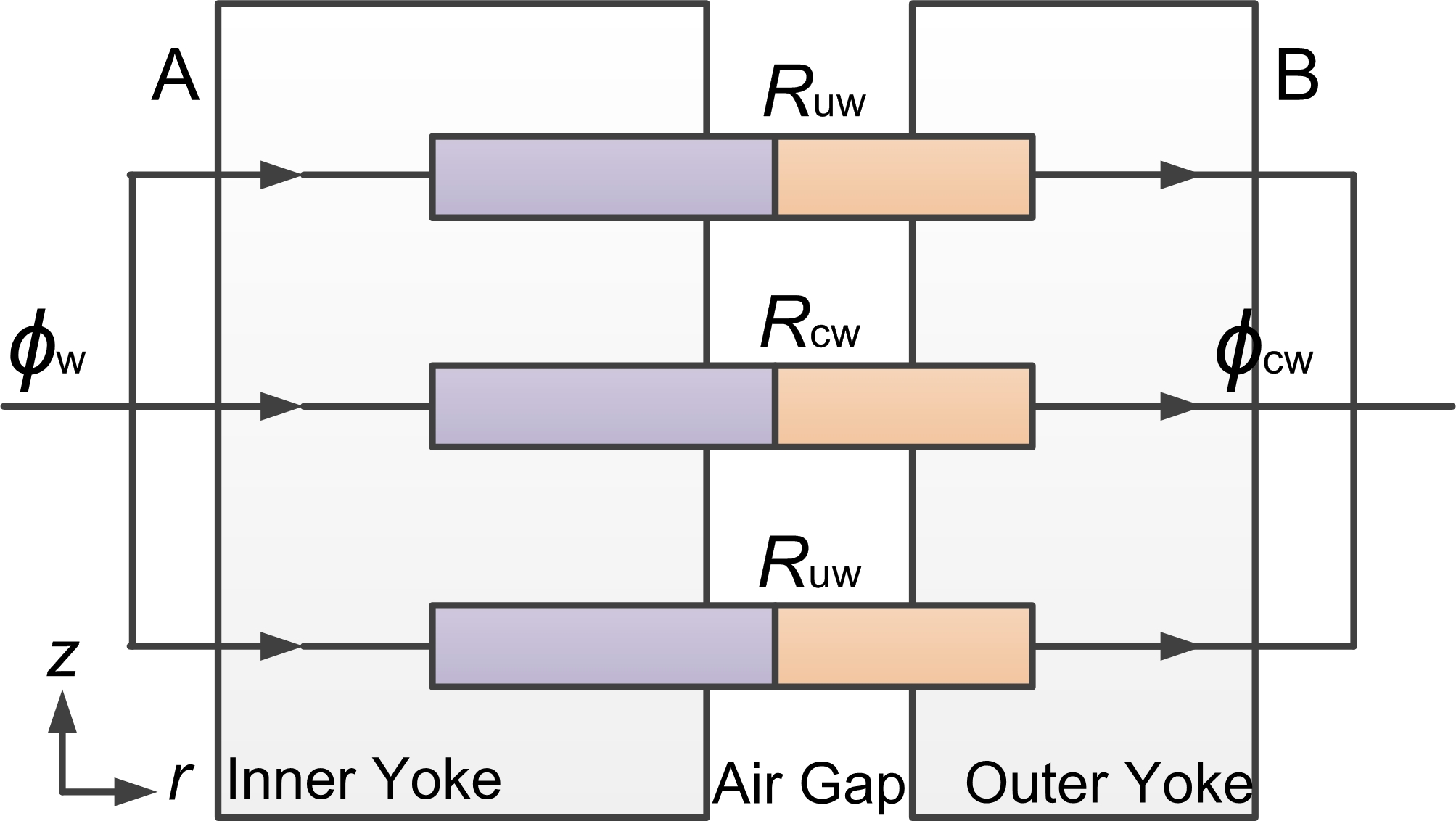}
\caption{Three-reluctance model of the magnet in weighing mode. Up-down symmetry about the center is assumed.}
\label{mag_reflux}
\end{figure}

\subsection{Redistribution of the magnetic flux density in the air gap due to the weighing current}

The weighing current in the coil produces an additional magnetic field which needs to be added to the already existing field produced by the permanent magnet system. The magnetic field produced by the magnet system in the yoke, near the gap, is in general uniform along the vertical axis. The magnetic field produced by the coil is largest at the coil position. Hence the reluctance of the yoke will change more at the coil position than above and below it. This nonuniform reluctance along the $z$ axis of the yoke will lead to a redistribution of the magnetic flux density in the gap. This redistribution causes the flux integral during the weighing mode, $(BL)_\mathrm{w}$, to be different from the flux integral during velocity mode, $(BL)_\mathrm{v}$.

Figure~\ref{mag_reflux} shows a simple model that can be used to evaluate this effect. A and B is a schematic representation of two vertical surfaces with the same magnetic potential, one in the inner yoke, the other in the outer yoke. The flux flows perpendicular through these two surfaces, such that in each measurement mode, the total flux through the two planes is considered to be identical. In the model, the magnet is divided in three vertical segments, the center segment (subscript $_{\mathrm{c}}$) contains the coil.  The model exhibits top-down symmetry, i.e. the upper segment is identical to the lower segment. $R_{\mathrm{c}}$ and $R_{\mathrm{u}}$  denote the reluctance of the center and the upper/lower segment, respectively. In velocity mode, the reluctances are the same, i.e., $R_{\mathrm{uv}}=R_{\mathrm{cv}}$.  Hence, the magnetic flux through each circuit is $\phi_{\mathrm{uv}}=\phi_{\mathrm{cv}}=\phi_\mathrm{v}/3$.

Two parts contribute to the reluctance of each segment: the reluctance of the air gap and that of the yoke. In weighing mode, the reluctances $R_{\mathrm{uw}}$ and $R_{\mathrm{cw}}$ can be written as
\begin{equation}
R_{\mathrm{uw}}=\frac{l_\mathrm{u}}{\mu_\mathrm{u} S_\mathrm{u}}+\frac{l_0}{\mu_0 S_\mathrm{u0}} ~\mathrm{and}~R_{\mathrm{cw}}=\frac{l_\mathrm{c}}{\mu_\mathrm{c} S_\mathrm{c}}+\frac{l_0}{\mu_0 S_\mathrm{c0}},
\end{equation}
where $l_\mathrm{u}$, $l_\mathrm{c}$ are the yoke lengths between surfaces A and B; $\mu_\mathrm{u}$ and $\mu_\mathrm{c}$ the permeability of the yoke for upper/lower and center segments. Note that three segments have the same geometrical parameters and the areas should be one third of the total, i.e.,  $l_\mathrm{u}=l_\mathrm{c}$,  $S_\mathrm{u}=S_\mathrm{c}=S_\mathrm{y}/3$, $S_\mathrm{u0}=S_\mathrm{c0}=S_\mathrm{0}/3$. The total flux divides according to the ratio of the inverse reluctances to that of the inverse of the total reluctance. The flux through the center circuit $\phi_{\mathrm{cw}}$ can be written as
\begin{equation}
\phi_{\mathrm{cw}}=\frac{1}{R_{\mathrm{cw}}}\left(\frac{2}{R_{\mathrm{uw}}}+\frac{1}{R_{\mathrm{cw}}}\right)^{-1}\phi_\mathrm{w}
=\left(2\frac{R_{\mathrm{cw}}}{R_{\mathrm{uw}}}+1\right)^{-1}\phi_\mathrm{w}.
\end{equation}
The relative change of the geometrical factor $BL$ at the weighing position (the center segment) in two modes is calculated as
\begin{equation}
\frac{(BL)_\mathrm{w}}{(BL)_\mathrm{v}}-1=
\frac{\phi_{\mathrm{cw}}}{\phi_{\mathrm{cv}}}-1
=\frac{\phi_\mathrm{w}}{\phi_\mathrm{v}}\frac{3}{2\frac{R_{\mathrm{cw}}}{R_{\mathrm{uw}}}+1}-1=(1+\xi_1)(1+\xi_2)-1\approx\xi_1+\xi_2.
\label{eq_xi}
\end{equation}
Here, $1+\xi_1=\phi_{\mathrm{w}}/\phi_{\mathrm{v}}$  and $1+\xi_2=3/(2R_{\mathrm{cw}}/R_{\mathrm{uw}}+1)$. An expression for $\xi_1$ is given in equation (\ref{eq_xi1F}), therefore only a calculation for $\xi_2$ is required. Similar to the discussion in section \ref{2.1}, $\xi_2$ is solved as
\begin{equation}
\xi_2=\frac{3}{2\frac{R_{\mathrm{cw}}}{R_{\mathrm{uw}}}+1}-1\approx\frac{2}{3}\left(1-\frac{R_{\mathrm{cw}}}{R_{\mathrm{uw}}}\right)\approx \frac{2S_0 l_\mathrm{c}\mu_0}{3 S_\mathrm{y} l_0\mu_\mathrm{v}} \left( 1-\frac{\mu_\mathrm{u}}{\mu_\mathrm{c}}\right).
\label{eq_xi2}
\end{equation}
Analogous to (\ref{eq:mup}), $\mu_\mathrm{c}$ and $\mu_\mathrm{u}$ can be obtained using
\begin{equation}
\mu_\mathrm{u}=\mu_\mathrm{v}+\frac{\displaystyle \left(\Delta H_\mathrm{u}\right)^2}{2 H_\mathrm{v}} \left. \frac{\partial \mu}{\partial H} \right|_{H=H_\mathrm{v}}
~\mathrm{and}~
\mu_\mathrm{c}=\mu_\mathrm{v}+\frac{\displaystyle \left(\Delta H_\mathrm{c}\right)^2}{2 H_\mathrm{v}} \left. \frac{\partial \mu}{\partial H} \right|_{H=H_\mathrm{v}},
\label{eq_muxi2}
\end{equation}
where $\Delta H_\mathrm{u}$ is the perpendicular magnetic field change in upper/lower segment and $\Delta H_\mathrm{c}$ is in the middle segment.

Substituting equation (\ref{eq_muxi2}) into (\ref{eq_xi2}) yields
\begin{equation}
\xi_2= \frac{2\mu_0 S_0 l_\mathrm{c}}{3 \mu_\mathrm{v}^2 S_\mathrm{y} l_0} \frac{\displaystyle \left(\Delta H_{\mathrm{c}}^2-\Delta H_{\mathrm{u}}^2\right)}{2 H_\mathrm{v}} \left. \frac{\partial \mu}{\partial H} \right|_{H=H_\mathrm{v}}.
\label{eq_xi2F}
\end{equation}

By adding $\xi_1$ in (\ref{eq_xi1F}) to $\xi_2$ in equation (\ref{eq_xi2F}) the total bias can be calculated as
\begin{equation}
\xi=\xi_1+\xi_2\approx\left(\frac{\frac{\displaystyle l_\perp}{\displaystyle S_\perp}}{\frac{\displaystyle l_0}{\displaystyle\mu_0 S_0}+\frac{\displaystyle l_\mathrm{m}}{\displaystyle\mu_\mathrm{m} S_\mathrm{m}}}+\frac{\frac{\displaystyle 2l_\mathrm{c}}{\displaystyle S_\mathrm{y}}}{ \frac{\displaystyle 3l_0}{\displaystyle\mu_0 S_0}}\left(\kappa_2^2-\kappa_1^2\right)\right)\frac{(\Delta H_\perp)^2}{2\mu_{\mathrm{v}}^2 H_\mathrm{v}} \left. \frac{\partial \mu}{\partial H} \right|_{H=H_\mathrm{v}},
\label{eq_xiF}
\end{equation}
where $\kappa_1=\Delta H_\mathrm{u}/\Delta H_\perp$ and $\kappa_2=\Delta H_\mathrm{c}/\Delta H_\perp$ are two magnetic field ratios.  As $\mu_{\mathrm{v\perp}}$ and $\mu_{\mathrm{v}}$ have similar values, it is reasonable to assume  $\mu_{\mathrm{v\perp}}\approx\mu_{\mathrm{v}}$.

The bias depends on the squared values of $\Delta H_{\perp}$, $\Delta H_\mathrm{c}$, $\Delta H_\mathrm{u}$ and hence quadratically on the current in the coil. Besides the current, the bias depends on parameters of the magnet system, most importantly at the working point of the yoke at $H=H_\mathrm{v}$. The bias can be eliminated by choosing parameters such that the yoke is at its maximum relative permeability, i.e.,
\begin{equation}
\left. \frac{\partial\mu}{\partial H}\right|_{H=H_\mathrm{v}}=0.
\end{equation}
To model magnet systems that differ from ideal systems described above, we introduce a new variable,
\begin{equation}
\delta=H_\mathrm{v}-H_\mathrm{m}\;\;\;\mbox{with}\;\;\; H_\mathrm{m} \;\;\mbox{such that}\;\;\left. \frac{\partial\mu}{\partial H}\right|_{H=H_\mathrm{m}}=0
\end{equation}
in next section.
\label{2.2}
\section{Evaluation and discussion}
In this section, the magnetic bias is calculated for typical parameters of a watt balance. To keep the analysis simple, we assume perfect up-down symmetry and that the position of the coil in the weighing mode is at the symmetry plane. Thus  an average magnetic field change in the yokes along the central horizontal axis $r$ could be used for calculating the $\Delta H_\perp$ value, i.e.,
\begin{equation}
\Delta H_\perp=\frac{\int_{l_{\perp}}\Delta H(r,z=0)dr}{l_{\perp}}.
\end{equation}
We further assume that in weighing mode the coil produces a force of $F=mg\approx 5$\,N, which is typical for a 1\,kg watt balance. In this case, the product of the coil current and the number of windings is given by a scalar form of the weighing equation as
\begin{equation}
NI=\frac{F}{2\pi r_0B_\mathrm{a}}=\frac{mg}{2\pi r_0B_\mathrm{a}},
\label{eq_NI}
\end{equation}
where $r_0$ is the mean radius of the coil and $B_a$ the mean value of the magnetic flux density at the coil position. The flux density contributed by the weighing current in the coil is calculated using the following approximations: The permeability of the yoke is set to the value at the working point, $\mu_\mathrm{v} = \mu(H_\mathrm{o})$ and the magnetomotive force of both magnets are set to zero. Since all flux produced by the coil flux in the yoke is perpendicular to the $r$ axis in the central plane $(z=0)$ and the additional magnetic density is continuous along the flux lines, the additional magnetic flux change in the yokes, $\Delta B_\perp=\mu_\mathrm{v}\Delta H_\perp$, can be considered to be equal to the flux in the yoke-air boundary. By Ampere's law, we have
\begin{equation}
2l_0\frac{\Delta B_\perp}{\mu_0}+l_\mathrm{y}\frac{\Delta B_\perp}{\mu_\mathrm{v}}=NI,
\label{eq_ampere}
\end{equation}
where $l_\mathrm{y}$ is the total length of the magnetic field  through the yoke and $l_0$ the width of the air gap. Since $\mu_\mathrm{v}>>\mu_0$, the second term can be neglected and $\Delta H_\perp$ is given by
\begin{equation}
\Delta H_\perp=\frac{\Delta B_\perp}{\mu_\mathrm{v}}=\frac{NI\mu_0}{2l_0\mu_\mathrm{v}}.
\label{eq_dHperp}
\end{equation}

\begin{figure}[!tp]
\centering
\includegraphics[width=3.5in]{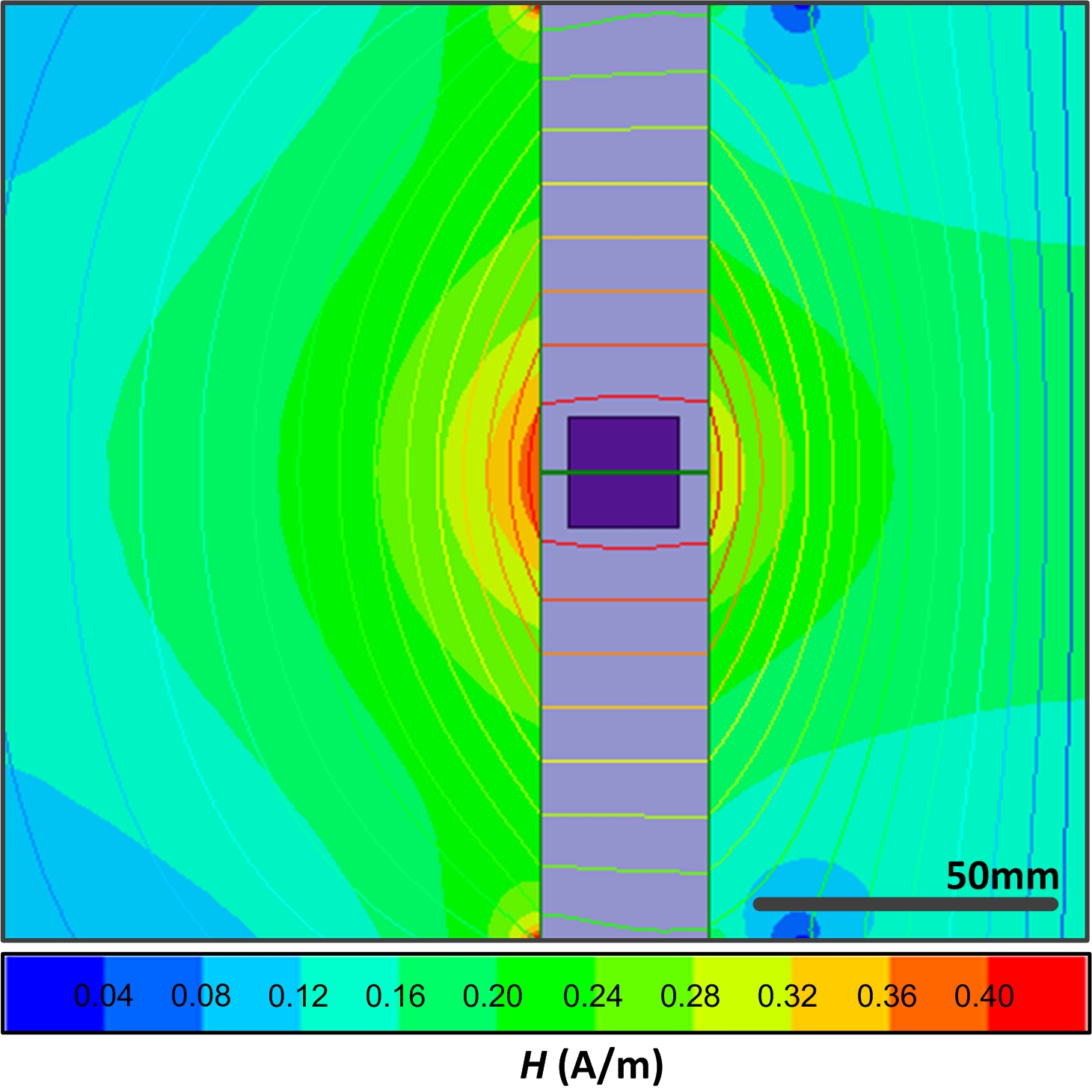}
\caption{FEM calculation of the magnetic field produced by the weighing current in the coil. For this calculation a gap width 3\,cm, a relative permeability of the yoke of 1000 and a magneto motive force of the coil of 8\,A\,turns is assumed.}
\label{mag_weighingpos}
\end{figure}

\begin{figure}[!tp]
\centering
\includegraphics[width=3.9in]{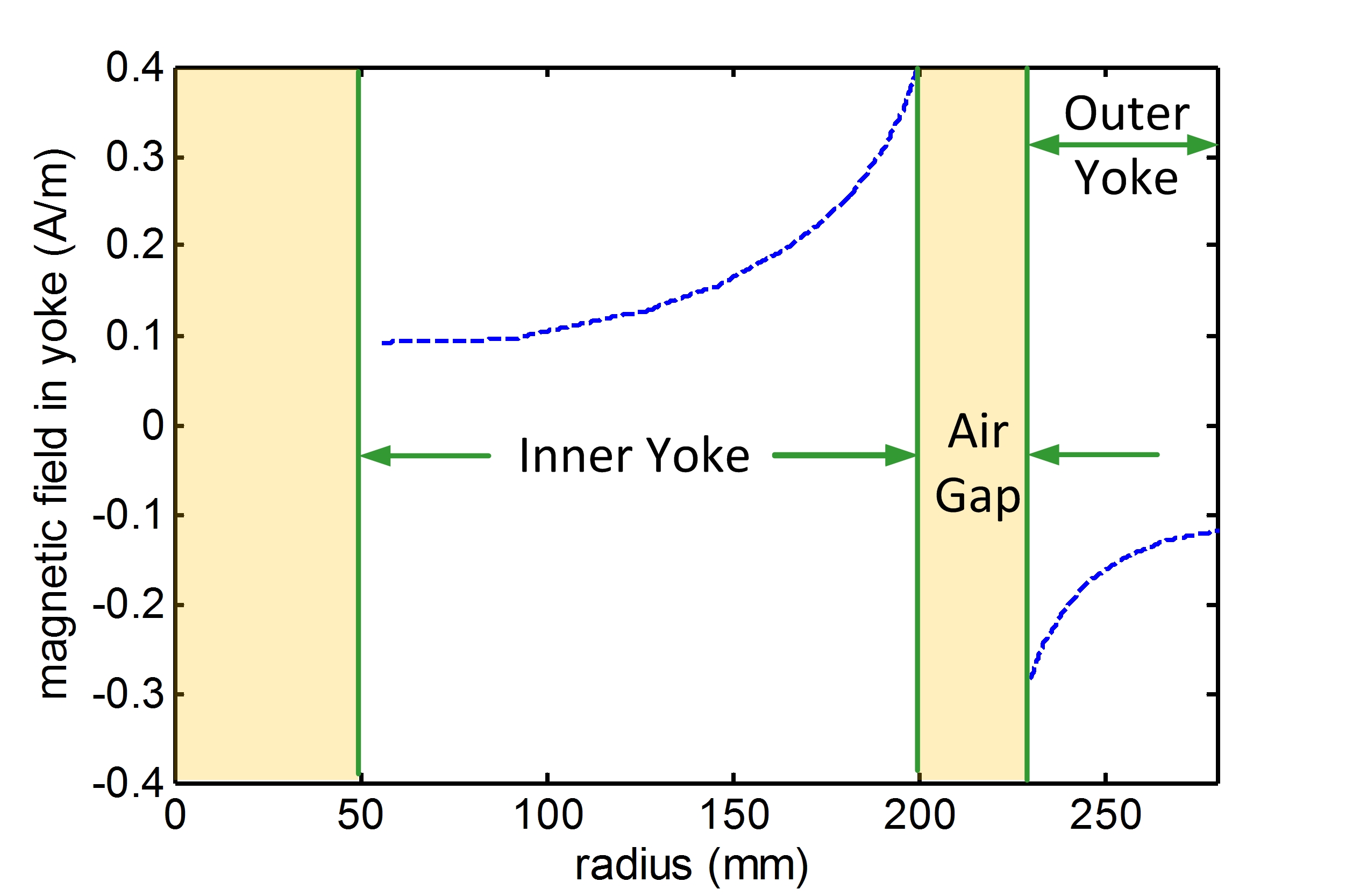}
\caption{The vertical magnetic field in yoke as a function of radial distance from the symmetry axis of the magnet. The same parameters as in figure~\ref{mag_weighingpos} were used for this FEM calculation.}
\label{mag_dHr}
\end{figure}

\begin{figure}[!tp]
\centering
\includegraphics[width=3.9in]{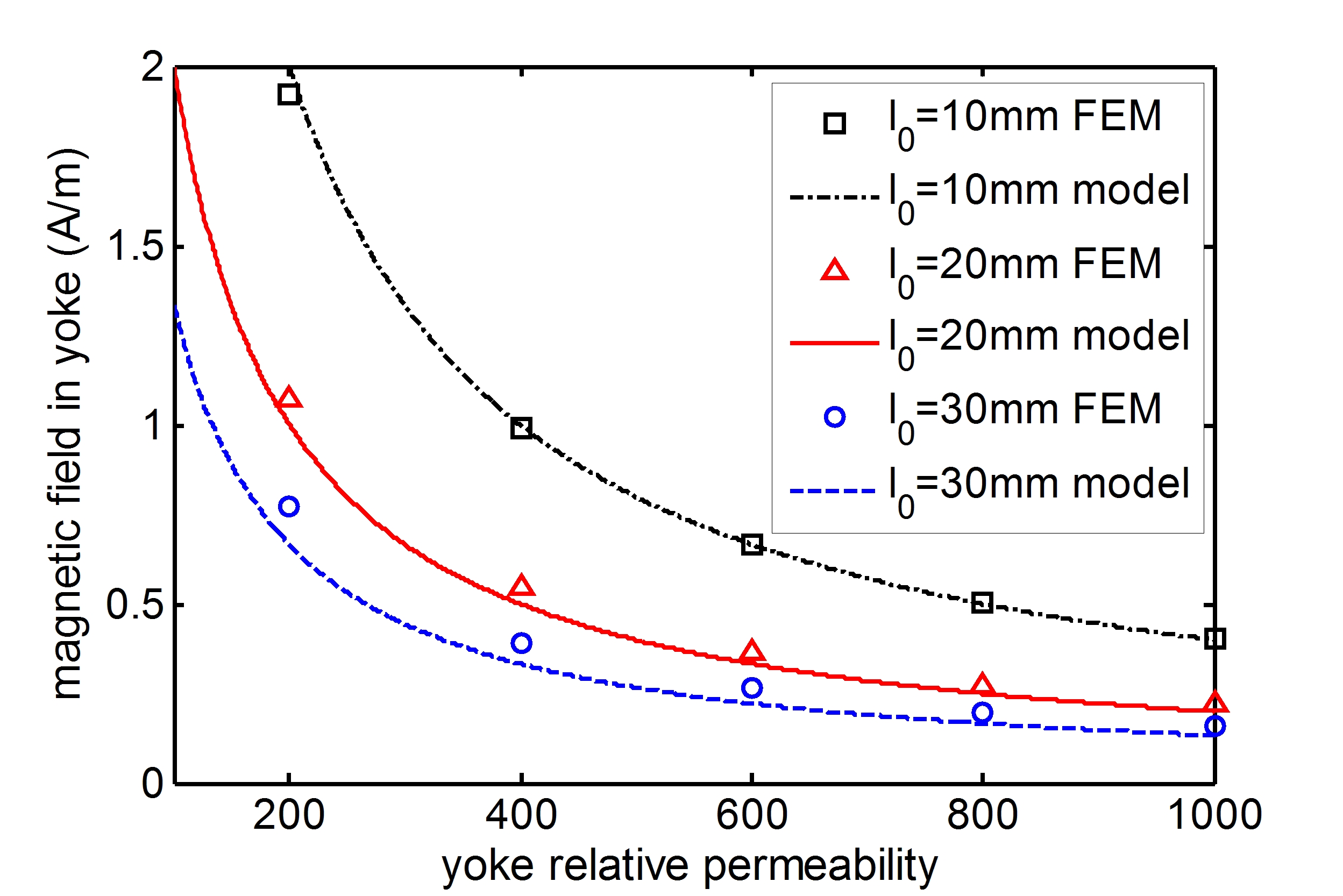}
\caption{Comparison of calculation results of $\Delta H_\perp$ by FEM and equation (\ref{eq_dHperp}) with different air gap widths and yoke permeabilities.}
\label{mag_equation}
\end{figure}

To verify equation (\ref{eq_dHperp}), calculations based on  the finite element method (FEM) were performed. For these  FEM calculations, an air gap width of $l_0=30$mm, a relative permeability of the yoke of $\mu_\mathrm{v}/\mu_0=1000$, and a magnetomotive force of the coil of $NI=8$\,A\,turns is assumed. Figure~\ref{mag_weighingpos} shows the magnetic field in an area around the coil. Figure~\ref{mag_dHr} shows the field in the plane of the coil as a function of radius. Both figures show that the magnetic field decreases rapidly with increasing distance from the coil. The FEM calculated mean magnetic field change in the yoke, i.e., $\Delta H_\perp$, is 0.16\,A/m which agrees with 0.13\;A/m calculated using the approximation (\ref{eq_dHperp}). FEM calculations with different yoke permeabilities and air gap widths were performed and compared to equation (\ref{eq_dHperp}), see figure \ref{mag_equation}.  The model agreed reasonably with the simulation for all 15 combinations. The agreement is better for smaller gap widths and larger relative permeabilities.

\begin{figure}
\centering
\includegraphics[width=3.9in]{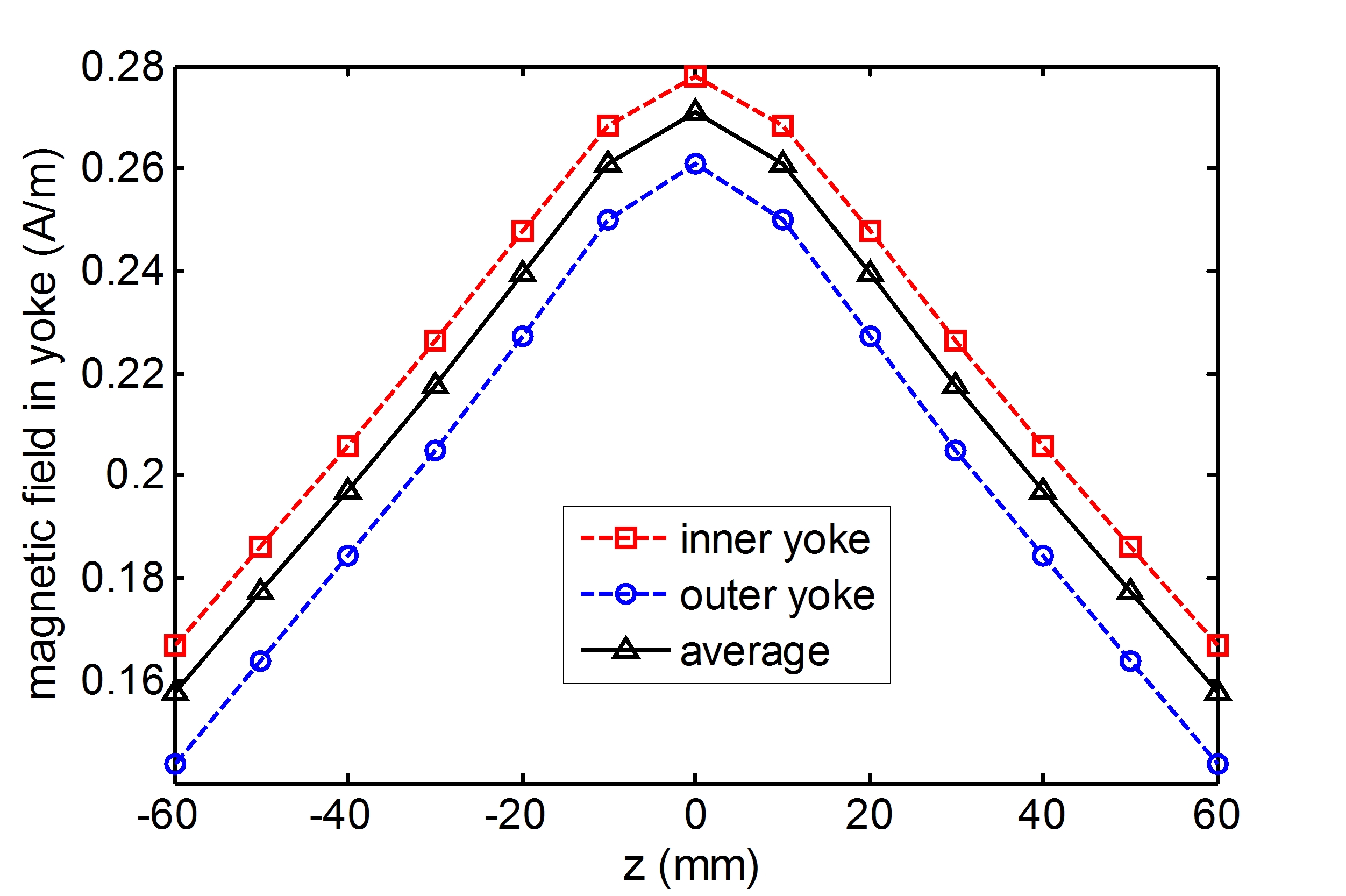}
\caption{The calculated perpendicular components of the magnetic field along the vertical direction. The average value is calculated with weights of 0.6 and 0.4 for the inner and outer yoke respectively.}
\label{mag_dHz}
\end{figure}

Substituting equation (\ref{eq_NI}) and equation (\ref{eq_dHperp}) into equation (\ref{eq_xi1F}), we obtain
\begin{equation}
\xi_1 \approx
\frac{\displaystyle l_\perp/\displaystyle S_\perp}{\displaystyle l_0/\displaystyle S_0+\displaystyle l_\mathrm{m}/\displaystyle S_\mathrm{m}}
\frac{m^2g^2\mu_0^3}{32\pi^2r_0^2B_\mathrm{a}^2l_0^2\mu_\mathrm{v}^3H_\mathrm{v}}
\left. \frac{\partial \mu}{\partial H} \right|_{H=H_\mathrm{v}}.
\label{eq_xi1expression}
\end{equation}

It can be seen from equation (\ref{eq_xiF}) that $\xi_2$ has a similar expression as $\xi_1$ and their ratio is only related to the magnet geometrical factor and a coefficient $\kappa_2^2-\kappa_1^2$, i.e., $\xi_2$ is solved as
\begin{equation}
\xi_2 \approx
\frac{\displaystyle 2l_\mathrm{c}/\displaystyle S_\mathrm{y}}{\displaystyle 3l_0/\displaystyle S_0}
\frac{m^2g^2\mu_0^3(\kappa_2^2-\kappa_1^2)}{32\pi^2r_0^2B_\mathrm{a}^2l_0^2\mu_\mathrm{v}^3H_\mathrm{v}}
\left. \frac{\partial \mu}{\partial H} \right|_{H=H_\mathrm{v}}.
\label{eq_xi2expression}
\end{equation}

In order to obtain the value of $\xi_2$, two magnetic field ratios $\kappa_1=\Delta H_\mathrm{u}/\Delta H_\perp$ and $\kappa_2=\Delta H_\mathrm{c}/\Delta H_\perp$ need to be calculated. Note that in equation (\ref{eq_xi2expression}) $\Delta H_\perp$, $\Delta H_\mathrm{c}$, and $\Delta H_\mathrm{u}$ are different integral quantities in the same magnetic field, hence both $\kappa_1$ and $\kappa_2$ are considered as constants. Here the two ratios are determined by FEM simulation with $l_0=30$mm, $\mu_\mathrm{v}/\mu_0=1000$. The distances between reference surfaces (A and B) and the air gap are 60mm and 40mm. The calculated perpendicular components of the magnetic field along the vertical axis $z$ are shown in figure \ref{mag_dHz}. It can be calculated from the simulation that $\kappa_1=0.16/0.16=1$ and $\kappa_2=0.27/0.16=1.7$.

Equations (\ref{eq_xi1expression}) and (\ref{eq_xi2expression}) determines the total bias $\xi$ as
\begin{equation}
\xi\approx
\left(\frac{\displaystyle l_\perp/\displaystyle S_\perp}{\displaystyle l_0/\displaystyle S_0+\displaystyle l_\mathrm{m}/\displaystyle S_\mathrm{m}}+(\kappa_2^2-\kappa_1^2)\frac{\displaystyle 2l_\mathrm{c}/\displaystyle S_\mathrm{y}}{\displaystyle 3l_0/\displaystyle S_0}\right)
\frac{m^2g^2\mu_0^3}{32\pi^2r_0^2B_\mathrm{a}^2l_0^2\mu_\mathrm{v}^3H_\mathrm{v}}
\left. \frac{\partial \mu}{\partial H} \right|_{H=H_\mathrm{v}}.
\label{eq_xiexpression}
\end{equation}

\begin{figure}
\centering
\includegraphics[width=3.9in]{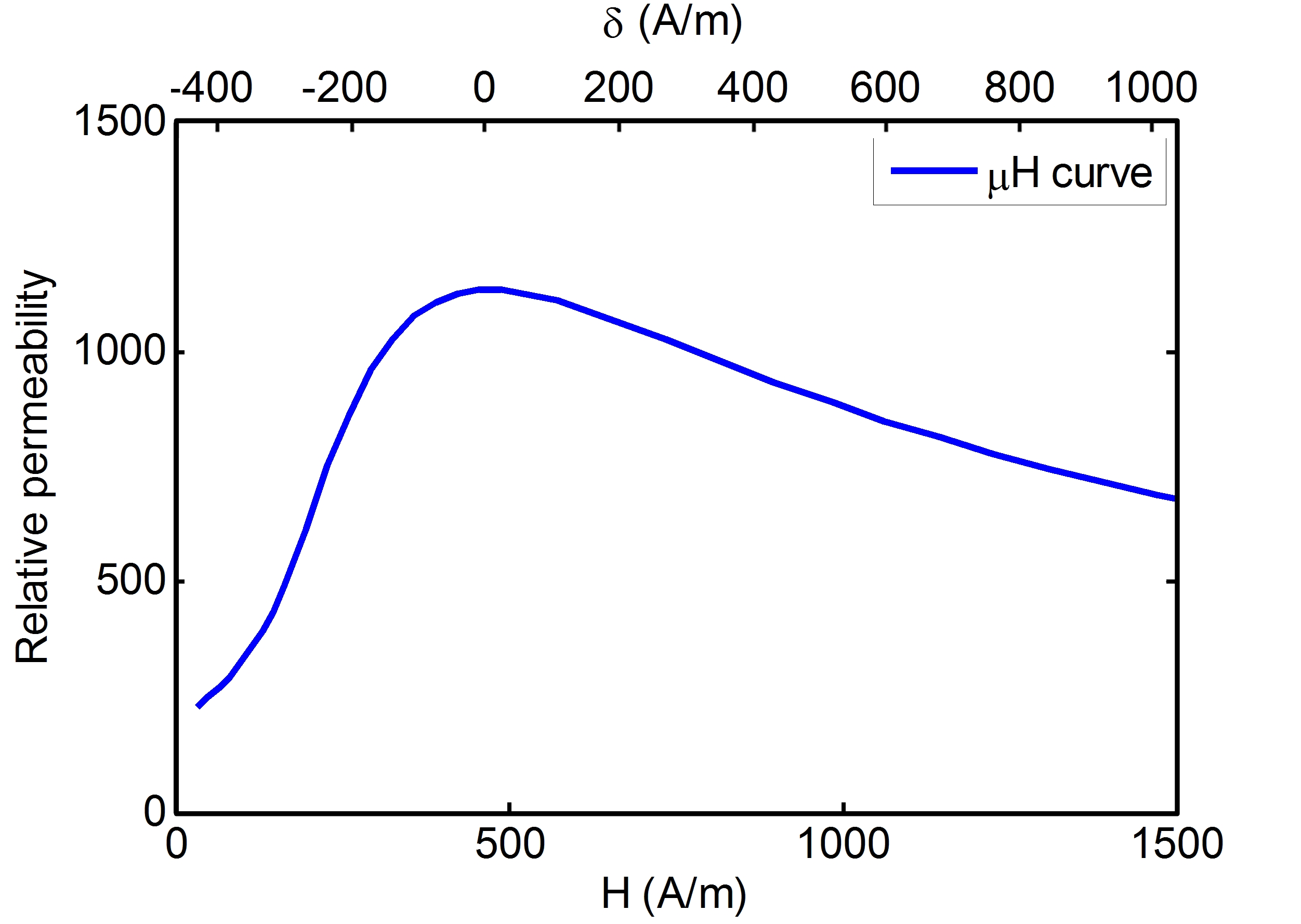}
\caption{The relative permeability as a function of magnetic field strength of AISI 1021 steel.}
\label{mag_muH}
\end{figure}
It can be seen from equation (\ref{eq_xiexpression}) that the bias is mainly related to three parameters: the magnetic flux density $B_\mathrm{a}$ in the air gap, the gap width $l_0$ and the dependence $\mu(H)$ of the yoke. In the evaluation, the $\mu H$ curve of AISI 1021 steel, which was used in  building the NIST-4 magnet, is assumed (shown in figure \ref{mag_muH}). The maximum relative permeability is 1137 at $H_\mathrm{m}=464$ A/m. Some geometrical factors are assumed as shown in table \ref{table1}.

\begin{table}[tb!]
\caption{Typical geometrical factors for a magnet system in a watt balance.}
\begin{indented}
\item[]
\begin{tabular}{@{}lll}
\br
Geometrical factor (ratio) &Value &Unit\\
\mr
$l_\mathrm{m}:l_\mathrm{y}:l_\parallel:l_\perp:l_\mathrm{c}$ & 50:300:100:200:100 &mm\\
$S_0:S_\mathrm{m}:S_\mathrm{y}:S_\parallel:S_\perp$ & 1:1:1:1:1 &~\\
$\mu_0:\mu_m$         & 1:1&~\\
$r_0$ &200&mm\\
\br
\end{tabular}
\end{indented}
\label{table1}
\end{table}

\begin{figure}[!tp]
\centering
\includegraphics[width=4.5in]{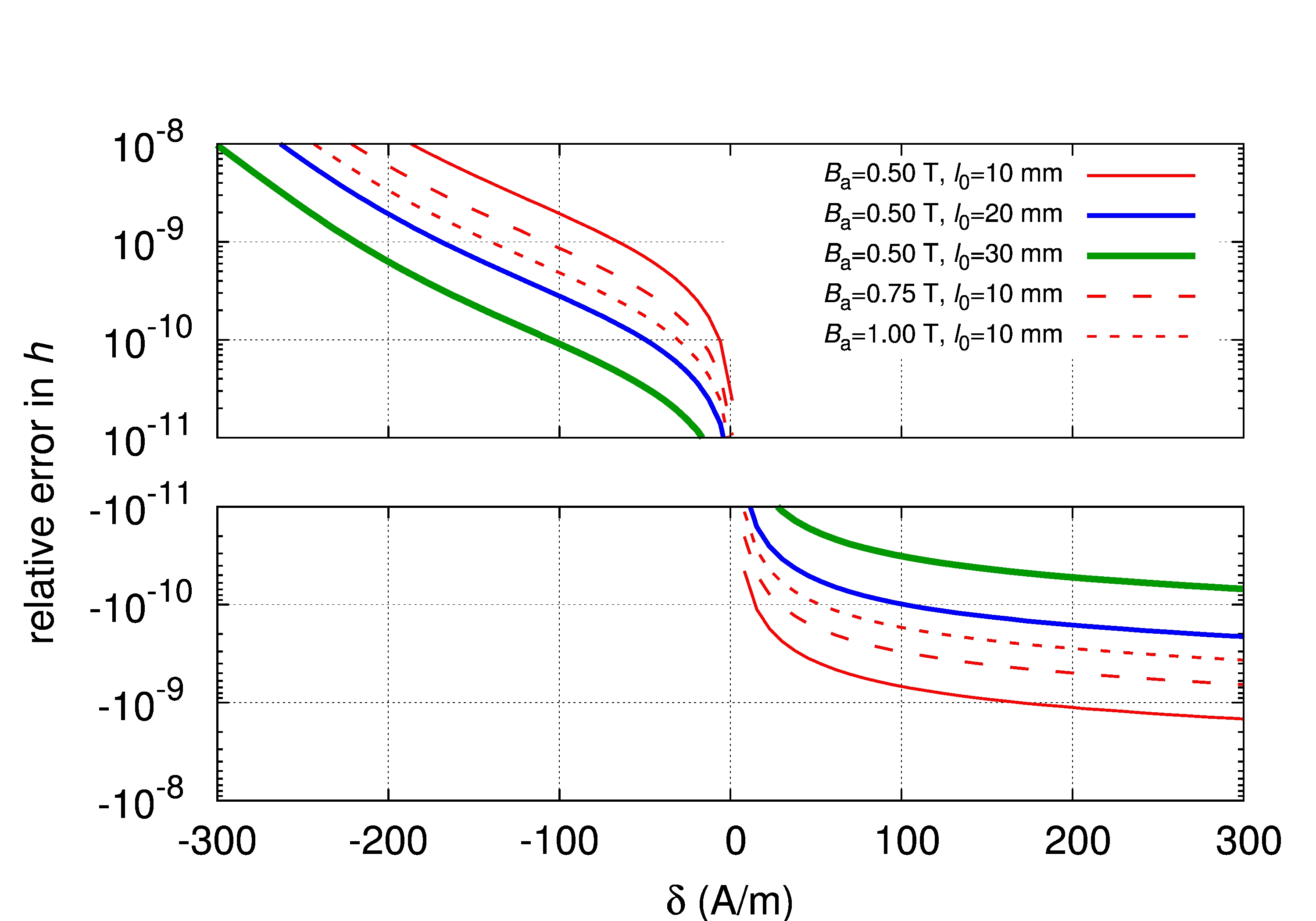}
\caption{Relative error for the Planck constant as a function of  $\delta$. Here $\delta=H_\mathrm{v}-H_\mathrm{m}$ with $H_\mathrm{m}$ such that $\left. \frac{\partial\mu}{\partial H}\right|_{H=H_\mathrm{m}}=0$.}
\label{mag_xi}
\end{figure}

In order to demonstrate the bias as a function of the magnetic field offset $\delta$, two different scenarios were considered. In the first scenario, the magnetic flux density in the gap remained the same $B_\mathrm{a}=0.5\,$T while the width of the air gap was changed. In the second scenario, the width remained the same $l_0=10$ mm and the flux density was changed. The results were expressed as the relative error of the Planck constant (the bias) as functions of the magnetic field strength offset $\delta$ and are shown in figure \ref{mag_xi}.

\begin{figure}[!tp]
\centering
\includegraphics[width=3.9in]{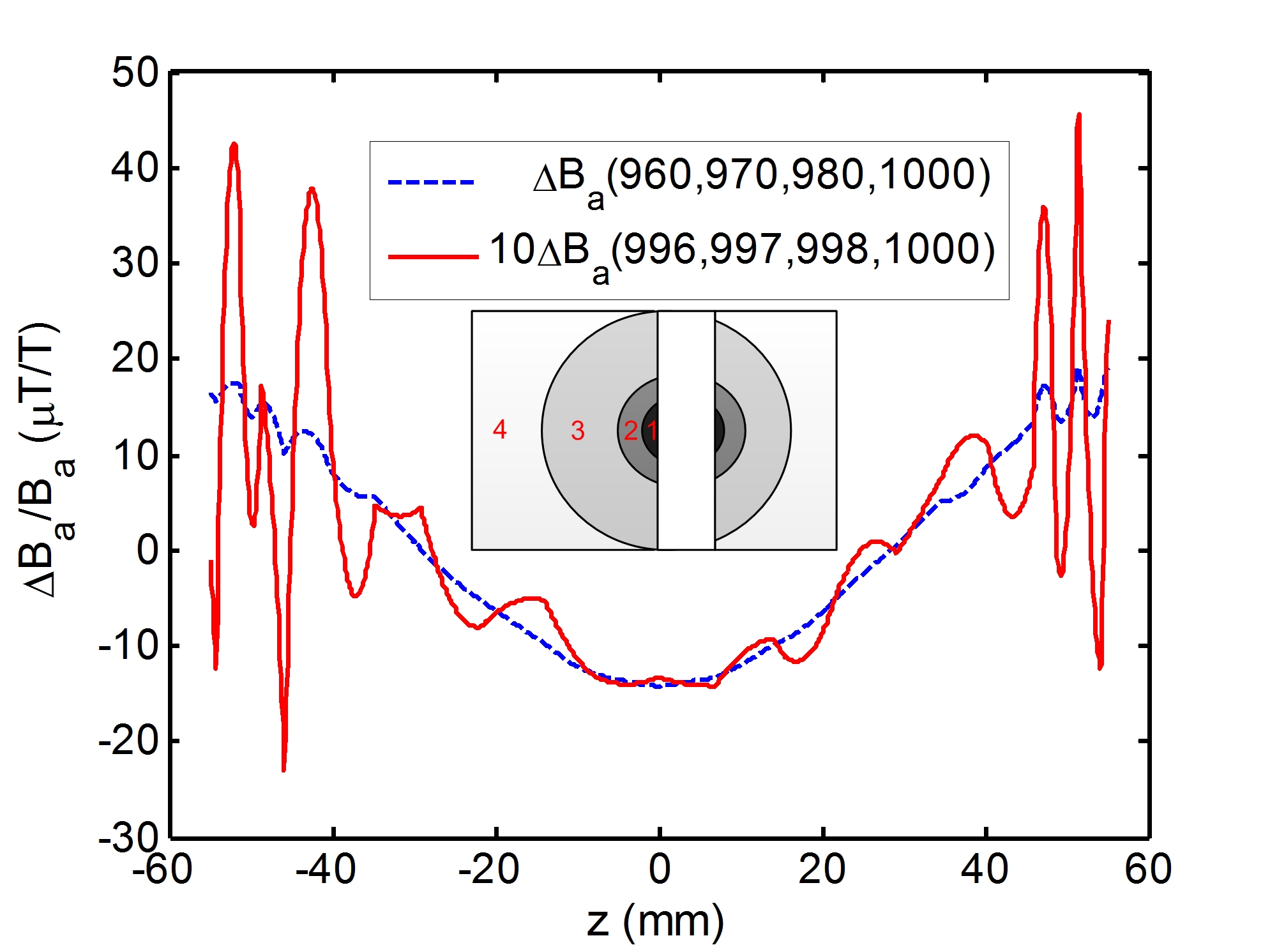}
\caption{Relative error for the Planck constant as a function of the magnetic field strength offset from the maximum permeability point.}
\label{mag_multi}
\end{figure}

As shown in figure \ref{mag_xi}, the bias has the opposite sign as the magnetic field offset $\delta$. Further, the slope of the bias for negative offsets is larger than for positive offsets.
Moreover, equation (\ref{eq_xiexpression}) shows that the bias  is (1) inverse proportional to $B_\mathrm{a}^2$; (2) inverse proportional to $\mu_\mathrm{v}^3$; (3) and depends critically on $l_0$ ( inverse to $l_0^n, 2<n<3$). A magnet design with a narrow  air gap benefits from a stronger magnetic field, but increases the bias error. In actual design of a watt balance, all parameters should be comprehensively optimized.

To verify the three-reluctance model for calculating $\xi_2$ in section \ref{2.2}, another FEM calculation was performed.  A multi-yoke structure at the weighing position is designed as shown in figure \ref{mag_multi} according to the coil flux contribution and all layers are set to different permeabilities where ($\mu_1,\mu_2,\mu_3,\mu_4$) denotes relative permeabilities of the yokes numbered 1, 2, 3, and 4. In order to obtain enough resolution, the contrast in permeabilities was exaggerated. The numbers (960, 970, 980, 1000) were used, which have a maximum difference in relative permeability of 40, about $4\times10^5$ larger than in reality. The simulation result is shown in figure \ref{mag_multi}. A second parameter set with (996, 997, 998, 1000), with  a maximum difference in relative permeability of 4 is also calculated. Its effect is about 10 times smaller than the first set. The result shows the nonlinearity is less than 7\%. Thus a relative change of the magnetic field at the weighing position can be estimated using $(15\times10^{-6})/(4\times10^5)= 3.8\times10^{-11}$ where the first value $15\times10^{-6}$ is read off the blue dashed line of figure \ref{mag_multi} at $z=0$ and the $4\times10^5$ is a scale factor assumed to scale the FEM simulation back to the range of permeability expected in reality. The FEM simulation agrees with the calculation result $4.3\times10^{-11}$ by equation (\ref{eq_xi2expression}).

Note, $\delta$ is not the average magnetic field difference of the whole yoke but the areas of the yoke adjacent to the coil in the  weighing position. In reality, $\delta$  can be quite large, e.g., several hundreds A/m. Table \ref{table2} gives a summary of the parameters (magnetic flux density in the air gap $B_\mathrm{a}$, air gap width $l_0$, and $\delta$ of watt balances built at different laboratories around the world. The paramter $\delta$ is calculated using the given value of $B_\mathrm{a}$, $l_0$, the mean radius of the air gap, and the  $BH$ curve of the AISI 1021 steel. The latter is a convenient assumption. In reality, different materials for yokes are employed. Hence, the numbers in the table are only an estimate.  The results show the bias amplitude from the magnetic nonlinearity is less than $1\times10^{-9}$, which is negligible with respect to the uncertainty goals of these watt balances.

\begin{table}[tb!]
\caption{Summary of $B_\mathrm{a}$, $l_0$ and calculation results of $\xi_1$ and $\xi_2$ for different watt balances. The $BH$ curve of the AISI 1021 steel is assumed.}
\begin{indented}
\item[]
\begin{tabular}{@{}lrrrrrr}
\br
Lab &$B_\mathrm{a}$ (T)	&$l_0$ (mm)& $\delta$ (A/m)& $\xi_1$ &$\xi_2$ \\
\mr
NPL-NRC	&0.45\cite{NRC2014}	&24\cite{NPL2007}  &-124	&$1.1\times10^{-10}$		&$2.1\times10^{-10}$	\\
LNE	   &0.95\cite{gournay05}	&9\cite{gournay05}   &270 	&$-1.1\times10^{-10}$	&$-4.5\times10^{-10}$		\\
BIPM	   &0.5\cite{stock06}	&13\cite{stock06}  &-97  	&$2.2\times10^{-10}$ &$6.6\times10^{-10}$	\\
METAS-2	&0.6\cite{baumann13}	&8\cite{baumann13}   &-38 	&$1.3\times10^{-10}$  	&$5.8\times10^{-10}$		\\
NIST-4	&0.55\cite{seiffert14}	&30\cite{schlamminger13}  &-69	&$1.6\times10^{-11}$ 	&$2.7\times10^{-11}$		\\
\br
\end{tabular}
\end{indented}
\label{table2}
\end{table}
All the evaluation and discussion are based on the analysis without considering the yoke $BH$ hysteresis. The hysteresis of the yoke may partly reduce this error, because the magnetic flux density in the weighing mode will remain for a while in the velocity mode. But the hysteresis effect, e.g., systematic effect from the non-symmetry of the minor $BH$ hysteresis loops, is complex and should be studied in the future. 

\section{Suggestions}
In this section, some suggestions are provided to reduce this nonlinear error.

The first conclusion is to make the working point for the yoke near weighing position approach the maximum permeability as much as possible, i.e., $\delta=0$. Based on equation (\ref{eq_xiexpression}), the best working point of the yoke near the weighing position is the zero crossing point of the error curve shown in figure \ref{mag_xi}. Note the working point for the yoke near weighing position here should be its mean value. As in the air gap, the magnetic flux density drops following a $1/r$ function ($r$ is the radius), the magnetic field for the inner yoke $H_\mathrm{in}$ is different from that of the outer yoke $H_\mathrm{out}$. From the calculation in figure \ref{mag_weighingpos}, a 50\% weight of magnetic field change can be applied for both inner and outer yokes, thus the design should meet
\begin{equation}
\frac{H_\mathrm{in}+H_\mathrm{out}}{2}=H_\mathrm{m}
\label{eq_design}
\end{equation}
To establish equation (\ref{eq_design}), an idea is to make adjustable magnetic compensations for the yoke around the weighing position. For example, current carrying compensation coils can be considered to generate opposite additional flux during the weighing mode. Also, small compensation permanent magnets can also shift the $BH$ working point of the yoke.

The second conclusion is that the bias error is inverse to the product $B_\mathrm{a}^2$, $\mu_\mathrm{v}^3$, and $l_0^n(2<n<3)$. Thus strong magnetic field $B_\mathrm{a}$, large air gap width $l_0$ and high permeability yoke are recommended for building a watt balance.

A third suggestion is to measure the amplitude of this effect in order to make possible corrections for the Planck constant value. The extrapolation method, known as the determination of the Planck constant by weighing different masses, has already realized by the watt balance community, but the measurement does not contain enough information when only two mass values are used. Measurements of the Planck constant with a greater variety of mass values may help to eliminate this error by extrapolating to the zero current value.

The magnetic error is caused due to the unsynchronized operating modes of the watt balance experiment. Thus investigations for synchronizing the weighing and velocity modes should be encouraged~\cite{fang13,robinson12b}.

\section{Conclusion}
A  nonlinear magnetic error in watt balance operation, which arises from the magnetic reluctance change of the yoke near the weighing position, is investigated. This error is proportional to the squared value of the  coil current. The analysis shows that this error can be optimized by making the yoke around the weighing position work at the maximum permeability point of the $BH$ curve. Further study evaluates the possible amplitude of the error as a function of the magnetic flux density difference between the actual and maximum-permeability points for the yoke near the weighing position. The result shows this nonlinearity is typically less than 1 part in $10^9$ which is negligible compared to a watt balance uncertainty of several parts in $10^{8}$. Therefore, at least in present stage, this nonlinear effect is not a limitation for watt balances.

\end{document}